# Superconductivity in the $Mo_2Re_3B$ compound

B. Andrzejewski[*]

Institute of Molecular Physics, Polish Academy of Science, Smoluchowskiego 17, PL-60179 Poznań, Poland

**Abstract.** We report on the synthesis and on basic superconducting properties of a completely new $Mo_2Re_3B$ ternary boride. The crystal structure of the $Mo_2Re_3B$ compound is characterised by Pmmm space group and the cell parameters: $a$=11.626 Å, $b$=8.465 Å and $c$=8.026 Å. The critical temperature is $T_c$=8.5 K, whereas the lower and the upper-critical fields at zero temperature are equal to $\mu_0H_{c1}(0)$=19.2 mT and to $\mu_0H_{c2}(0)$=3.7 T, respectively. The corresponding Ginzburg-Landau parameter is equal to $\kappa$=16.5 and the superconducting gap is estimated to be $2\Delta/k_BT_c \approx 3.2$.

*Keywords*: Superconductors (A); Critical field (C); critical temperature (C); Magnetic measurements (D)

## 1. Introduction

Recent discovery of superconductivity in magnesium diboride $MgB_2$ [1] has also renewed substantial interest to other boride intermetallics. For many years, these compounds were regarded as promising candidates for "high-temperature" superconductors due to high frequency vibrations of lighter elements and strong electron-phonon interactions, which should facilitate enhanced critical temperatures. From this point of view the best superconducting systems are $AB_2$-type diborides containing light $A$ alkali metal like Li, Be or

---
[*] Tel.: +48 61 8695 283; Fax: +48 61 8684 524. *E-mail address*: and@ifmpan.poznan.pl

Al. Unfortunately, in stoichiometric diboride compounds superconductivity is usually absent. This is in contrast to nonstoichiometric phases, like $MoB_{2.5}$, $NbB_{2.5}$, $Mo_2B$, $W_2B$ and $BeB_{2.75}$ [2-5], which exhibit superconductivity.

According to some arguments, superconductivity can be induced in large clusters of light atoms. Indeed, some compounds containing 3D clusters of metals, like alkali metal doped fullerides $M_3C_{60}$ ($M$=K, Na, Rb, Cs) exhibit enhanced critical temperatures, including the record transition temperature $T_c$ equal to 33 K, observed in the $RbCs_2C_{60}$ fulleride. Atoms of boron possess similar natural tendency to form clusters. Examples are octahedral $B_6$ clusters in $MB_6$ compounds, icosahedral $B_{12}$ clusters in $\beta$-rhombohedral boron and cubo-octahedral $B_{12}$ clusters in $MB_{12}$ intermetallics. Several cubic hexaborides $MB_6$ and dodecaborides $MB_{12}$ with $M$=Sc, Y, Zr, La, Lu, Th turn out to be superconducting [6,7], whereas this type borides with $M$=Ce, Pr, Nd, Eu, Gd, Tb, Dy, Ho, Er, Tm are ferromagnetic or antiferromagnetic [8,9]. There is even reported, that the superconductivity in $YB_6$ ($T_c$=6.5÷7.1 K) and in $ZrB_{12}$ ($T_c$=6.03 K) compounds appears exactly due to the clusters of light boron atoms [3].

Unfortunately, many reports on superconducting properties of boron compounds are contradictory. This concerns for example $NbB_2$ and $TaB_2$ diborides. The first compound is reported to be a normal-metal [2] or superconductor at $T_c$=0.62 K [10] or at $T_c$=3K [11]. Samples of the second boride exhibit normal metal [10] or superconducting behaviour with relatively high critical temperature equal to $T_c$=9.5 K [12]. There are also known superconducting borides with rhenium like: $Re_2B$ ($T_c$=2.8 K) [13], $Re_3B$ ($T_c$=4.7 K), $ReB_2$ ($T_c$=4.5÷6.3 K) [14], $Re_7B_3$ ($T_c$=3.3 K) [15].

Besides simple borides, there have been recently successfully synthesized and studied $Mo_7Re_{13}B$ and $W_7Re_{13}B$ ternary borides [16,17]. These compounds exhibit rather complex crystallographic structures and superconducting properties. They are extreme type-II superconductors with the Ginzburg–Landau parameters $\kappa$≈54 and $\kappa$≈101 for $W_7Re_{13}B$ and for $Mo_7Re_{13}B$, respectively. In the case of the $W_7Re_{13}B$ compound there is even present pronouncedt irreversibility line and large reversible part of H-T diagram. The transition temperatures for $W_7Re_{13}B$ and for $Mo_7Re_{13}B$ superconductors are rather low and equal to $T_c$=7.1 K and to $T_c$=8.3 K, respectively. The lower- and upper-critical fields exhibit large spread in theirs values: for $W_7Re_{13}B$ they are equal to $\mu_0H_{c1}(0)$=7.7 mT, and to $\mu_0H_{c2}(0)\approx$11.4 T, whereas for the $Mo_7Re_{13}B$ compound $\mu_0H_{c1}(0)$=3.0 mT only, and $\mu_0H_{c2}(0)$=15.4 T. The upper-critical fields are rather high as for conventional superconductors.

Recently, we have reported on superconductivity in the new $Mo_2Re_3B_x$-$Mo_3Re_2B_x$ (where x≈1) eutectic system composed of two, previously unreported superconducting phases [18,19,20]. The present study is devoted to the problem of synthesis and to the basic superconducting properties of these phases, namely to the $Mo_2Re_3B$ compound.

## 2. Experimental

The $Mo_2Re_3B$ compound was synthesized using the method of induction melting. Stoichiometric amounts of the starting materials: Mo and Re fine powders were mixed with amorphous B powder and alloyed under an argon atmosphere in a water-cooled boat. After alloying, the resulting ingot with the total mass of about 0.5 g, was inverted and remelted several times to assure good homogeneity of the sample. The ingot was next cut into three samples for magnetic, electric transport and X-ray diffraction measurements, respectively.

The magnetometric and electric transport measurements were performed by means of Oxford Instruments Ltd. MagLab 2000 System AC susceptometer/DC magnetometer equipped additionally with the electric transport probe. A standard four point contact method was applied in the case of electric measurements. The demagnetizing co-efficient of the sample used in magnetometric measurements was assumed to be $N\approx$1/3. The magnetometric measurements were performed applying both zero field cooling (ZFC) and field cooling (FC) procedures. In ZFC procedure, a sample was cooled to the low temperature in absence of a magnetic field. Next the field was applied and the measurements were taken when warming of the sample. During FC procedure the sample was cooled in presence of the magnetic field and the measurements were simultaneously performed.

The structure of this compound was examined by means of X-ray diffraction. The X-ray patterns were taken at room temperature using CuKα radiation. Next, the diffraction patterns were analyzed by means of Retveld method using FULLPROF package.

## 3. Results and discussion

The X-ray diffraction pattern of the compound with nominal composition $Mo_2Re_3B$ is presented in Fig. 1. In general, the pattern indicates presence of well crystallized compound in the examined sample. The solid line is the best fit of the theoretical spectrum obtained in Retveld analysis to the experimental data represented by full points. The vertical bars below the diffraction pattern represent the Bragg peak positions for orthorhombic Pmmm structure. The solid line at the bottom of Fig. 1 show the deviation between the data and fitted spectrum. The refined crystallographic cell parameters are as follows: $a$=11.626 Å; $b$=8.465 Å and $c$=8.026 Å. The exact crystallographic structure of this compound is yet unknown and will be a subject for future studies.

The superconductivity of the $Mo_2Re_3B$ compound was confirmed by means of magnetic and electric resistivity measurements presented in Fig. 2. In magnetic measurements, the onset of diamagnetic signal is observed below the critical temperature equal to 8.5 K. The onset of superconductivity is very narrow with the transition width at 50% of diamagnetic moment equal to 0.1 K. This indicates good purity of the sample. The ZFC and FC magnetization curves differ much in magnitude despite the applied magnetic field was low and equal to 1 mT, only. This behaviour proves substantial flux pinning in FC mode, even in the range of low magnetic fields.

In contrast to the magnetometric data, the resistive superconducting transition, shown in the inset to Fig. 2, is rather wide. Moreover, the zero resistivity state is never attained, even at the lowest temperature available equal to about 3 K. The absolute value of resistivity at room temperature equal to ≈90 µΩcm, is rather high, similar to the high value of the residual resistivity ratio defined as $RRR=\rho(T \approx T_c)/\rho(T=300K)$ and equal to 0.55.

The discrepancy between magnetic and resistive superconducting transition is probably related to the granular structure of the sample. Furthermore, many authors reported on the boron excess or on the presence of boron impurities deposited on grain boundaries which leads to the increased or even nonmetallic behaviour of the resistivity in these boride compounds [11, 16, 17, 20, 21]. Thus the magnetic measurements monitor the superconducting transition inside superconducting grains whereas the resistive transition is governed by the process of percolation in this granular system. The percolation begins immidiately below the onset of superconductivity inside the grains, however at high temperature majority of grains is still decoupled. This means that the phase of the wave function is strongly affected by thermal fluctuations and evolutes independent in each of the grain. In low temperatures some of the grains become coupled or, in other words, phase-locked and start to form clusters. When the temperature further decreases, the clusters expand and connect each other. Below the decoupling temperature $T_{cJ}$ clusters covers whole sample and the resistivity vanish. The decoupling temperature is inversely proportional to the intergranular critical current density $J_c(0)$ at zero temperature [22]; $T_c - T_{cJ} \propto T_c^2 / J_c(0)$, where $T_c$ is a critical temperature of grains. In the case of $Mo_2Re_3B$ compound the percolation is never completed and the sample always exhibits some normal resistivity. This means that the decoupling temperature is below 3 K and there is no circulation of intergranular supercurrents in the total volume of the sample, however some small superconducting clusters may already exist.

A set of magnetization curves $M(T)$ recorded for various applied magnetic fields is shown in Fig. 4. The superconducting transition manifests itself as an onset of diamagnetic signal, which takes place in the so-called onset-critical temperature $T_c^{onset}$. Under the influence of a magnetic field the transitions broaden and become more gradual. The exact values of the critical temperatures corresponding to given magnetic fields $T_c^{onset}(H)$ may be transformed into the equivalent dependence of the upper-critical field on temperature $H_{c2}(T)$, which is presented in the inset to Fig. 4. It is evident that the temperature dependence of the upper-critical field $H_{c2}(T)$ exhibits negative curvature, typical for conventional superconductors. At higher temperatures the upper-critical field is almost linear with respect to temperature and the slope $dH_{c2}/dT$ is equal to -0.884 T/K.

The latter co-efficient allows one to evaluate the upper critical field $H_{c2}(0)$ at zero temperature. In the framework of the Werthamer-Helfand-Hohenberg (WHH) theory [23, 24] $H_{c2}(0)$ is given by the formula; $H_{c2}(0) \cong -0.69 T_c (dH_{c2}/dT)$. Using the above determined value of the slope and the critical temperature $T_c$=8.5 K one obtains the overestimated result $\mu_0 H_{c2}(0)$=5.2 T, which is above the experimental data. However, it was argued, that for some compounds, for example $MgCNi_3$, the factor 0.69 should be replaced by lower value [25] for example by 0.59. The latter co-efficient gives here more realistic result $\mu_0 H_{c2}(0)$=4.4 T. In any case, the upper critical field is well below the paramagnetic limit $\mu_0 H_{cP}=1.84 T_c$ and equal to 15.6 T. This proves that the Zeeman coupling only slightly influences the superconducting properties of the $Mo_2Re_3B$ compound. If the effects of scattering are negligible and the superconductor belongs to the clean limit, the experimental $H_{c2}(T)$ data can be fitted with the analytical expressions derived by Maki [26]:

$$\mu_0 H_{c2}(t) = c_1 (1+\alpha^2)^{-1/2} \left(1 - \frac{2}{3}\gamma^2 t^2 \frac{1-\alpha^2}{1+\alpha^2}\right) \qquad c_1 = \frac{3\Delta(0)}{2e\tau_{tr}v^2} \qquad t \ll 1 \qquad (1)$$

$$\mu_0 H_{c2}(t) = c_2 \theta \left[1 - \theta\left(\frac{1}{2} - \frac{28\zeta(3)}{\pi^4}(1-\alpha^2)\right)\right] \qquad c_2 = \frac{12k_B T_c}{\pi e \tau_{tr} v^2} \qquad t \leq 1 \qquad (2)$$

Where $t=T/T_c$ is the reduced temperature, $\theta=1-t$, $\gamma=1.78$, $\zeta(3)=1.202...$ is the Riemann zeta function. The Maki co-efficient $\alpha$ is given by $\alpha = \sqrt{2}H'_{c2}(0)/H_{cP}$, where $H'_{c2}(0)$ is the upper critical field in the absence of the Pauli term. The symbols in the $c_1$ and $c_2$ co-efficients are as follows: $v$ is the Fermi velocity, $\tau_{tr}$ is the collision time and $\Delta(0)$ is the superconducting gap at zero temperature.

The best fit to the experimental data $H_{c2}(T)$ is obtained for $T_c=8.45$ K and for $\alpha=0.6$, whereas the theoretical value of $\alpha$, calculated from the WHH formula is equal to 0.47. From the fit, the upper critical field at zero temperature is equal to $\mu_0 H_{c2}(0)=3.7$ T. On the other hand, the ratio $c_1/c_2=\pi\Delta/8k_B T_c$ allows one to roughly evaluate the basic BCS parameter $2\Delta/k_B T_c$. From the best fit; $c_1/c_2=0.62$ which gives the ratio $2\Delta/k_B T_c \approx 3.2$. This value corresponds to the weak coupling BCS limit with the superconducting gap equal to $\Delta \approx 1.2$ mV.

The lower critical field $H_{c1}(T)$ can be fitted in similar manner i.e. using the relation between $H_{c1}(t)$ and $H_{c2}(t)$:

$$H_{c1}(t) = H_{c2}(t) \frac{\ln \kappa(t)}{\kappa^2(t)} \qquad (3)$$

Where the Ginzburg-Landau parameter $\kappa(t)$ is now temperature dependent and expressed as:

$$\kappa(t) = 1.2 \frac{\kappa(0)}{\sqrt{1+\alpha^2}} \left(1 - 1.05 \frac{1-2\alpha^2}{1+\alpha^2} t^2\right) \qquad t \ll 1 \qquad (4)$$

$$\kappa(t) = \kappa(0)\left(1 + (0.119 - 0.346\alpha^2)\theta\right) \qquad t \leq 1 \qquad (5)$$

The experimental values of the lower critical field at selected temperatures can be found using as a criterion the deviation of initial part of magnetization $M(H)$ from linearity. This procedure is demonstrated in the inset to Fig. 5. The data determined in this way should be next corrected for the demagnetizing effect. From Fig. 5 it is clear that, the lower critical field once again exhibits usual, negative curvature common for conventional superconductors.

The best fit of the eq. (3) to the experimental $H_{c1}(T)$ data presented in Fig. 5 is obtained for $\kappa(0)=16.5$, where $\kappa(0)$ is the only one adjustable parameter. Others parameters in eqs. (4) and (5), namely the critical temperature $T_c$ and the Maki parameter $\alpha$ are identical to those obtained earlier, from $H_{c2}(T)$ fitting, and here treated as constants. The zero-temperature value of the lower critical field determined from the best fit is equal to $\mu_0 H_{c1}(0)=19.2$ mT.

From the values of the critical fields it is now possible to evaluate other basic parameters of this superconductor like the coherence length, the penetration depth and the Ginzburg-landau co-efficient. The coherence length $\xi$ can be calculated from the relationship; $\mu_0 H_{c2}=\Phi_0/2\pi\xi^2$ where $\Phi_0 = 2\cdot10^{-15}$ Wb is flux quantum. For $\mu_0 H_{c2}=3.7$ T it is equal to 94 Å. Similarly, on the basis of the relation; $\mu_0 H_{c1}(0)= (\Phi_0/4\pi\lambda^2)\ln(\lambda/\xi)$ one can determine the penetration depth $\lambda=1550$ Å. The Ginzburg-Landau co-efficient $\kappa$ determined from the definition $\kappa=\lambda/\xi$ is equal to $\kappa=16.5$, which is in good agreement with the value obtained directly from the fit.

Knowing the penetration depth $\lambda$, the coherence length $\xi$ and the resistivity of the sample $\rho$ one can evaluate now the mean free path $l_e$ and verify whether the superconductor belongs to the clean ($\xi<<l_e$) or to the dirty limit ($\xi>>l_e$). Using the relations in the clean limit: $\xi=0.74\xi_0$, $\lambda^2=\lambda_L^2/2$ where: $\xi_0=0.18\hbar v/k_B T_c$, $\lambda_L^2=m/\mu_0 n e^2$ and the Drude model: $\rho_i=mv/ne^2 l_e$ it is easy to derive the following formula: $l_e=15\mu_0\lambda^2\xi k_B T_c/\hbar\rho_i$. For the experimental data one obtains the mean free path $l_e\approx 90$Å, which is comparable to the coherence length $\xi$. However, because of granularity and an interfacial boron in the $Mo_2Re_3B$ sample, the experimental value of resistivity $\rho$ used here for the calculations, may be even several times larger than the intrinsic one $\rho_i$. Thus, the actual value of $l_e$ is much larger than $\xi$ and the superconductor belongs to the clean limit.

Figure 6 presents a set of magnetization loops $M(H)$ recorded for the $Mo_2Re_3B$ compound at various temperatures. The magnetization loops are asterisk-like, which indicates that critical current density is strongly suppressed by the magnetic field. The substantial pinning and critical current exists in the low magnetic field range, only. Above some magnetic field, the parts of loop recorded for increasing and for decreasing magnetic field merge each other and the magnetization becomes totally reversible. These characteristic fields correspond exactly to the values of the upper critical field $H_{c2}(T)$ at given temperatures. This proves that there is no irreversibility line in the $Mo_2Re_3B$ compound, or the irreversibility line is very close to $H_{c2}(T)$. This is in agreement to many reports concerning conventional superconductors for which no irreversibility lines were found. However, in some ternary borides, for example in the $W_7Re_{13}B$ compound, the irreversibility line exists and is well separated from the upper critical field [22] making large part of the H-T diagram totally reversible.

## 4. Conclusions

We have synthesized for the first time, superconducting $Mo_2Re_3B$ ternary boride, with the transition temperature equal to $T_c$=8.5 K. This compound is characterised by Pmmm symmetry with the crystallographic cell parameters equal to: $a$=11.626 Å; $b$=8.465 Å, $c$=8.026 Å. It is granular superconductor because the metallic $Mo_2Re_3B$ grains are separated by thin layer of excess boron which leads to the enhanced resistivity.

This superconductor belongs to the weak BCS coupling clean limit with the parameter $2\Delta/k_BT_c$ close to 3.2. It also exhibits type II superconductivity characterised by moderately high Ginzburg-Landau parameter $\kappa$=16.5 and irreversible properties with no trace of irreversibility line in H-T diagram. The temperature dependence of the lower $H_{c1}(T)$ and the upper critical field $H_{c2}(T)$ can be well described within the Maki theory. Both critical fields exhibit negative curvature, common for conventional superconductors. The zero-temperature values of the critical fields are equal to: $\mu_0 H_{c1}$=19.2 mT and $\mu_0 H_{c2}$=3.7 T, which corresponds to the penetration depth $\lambda$=94 Å, to the coherence length $\xi$=1550 Å

**Acknowledgements** B.A. wish to thank Dr. A. Szlaferek for preparation of the $Mo_2Re_3B$ sample and Dr. T. Toliński for discussion concerning the X-ray diffraction data.

**Figure Captions**

Fig. 1 The XRD pattern of the $Mo_2Re_3B$ compound. The vertical bars indicate positions of XRD peaks for orthorhombic Pmmm structure. The solid line at the bottom of the figure shows the difference between the experimental XRD data and the fit.

Fig. 2 The magnetic superconducting transition for the ZFC and FC procedure and applied magnetic field equal to 1 mT (main panel). The resistive transition is shown in the inset.

Fig.3 The magnetic superconducting transitions for various applied magnetic fields (main panel). The inset shows the upper critical field $H_{c2}(T)$ dependence on the temperature. The solid line is the fit according to the Maki model.

Fig.4 The magnetization dependence $M(H)$ on applied magnetic field for two selected temperatures (main panel). For clarity, not all data are shown in this plot. The inset presents lower critical field $H_{c1}(T)$ dependence on temperature. The solid line is the fit according to the Maki model.

Fig. 5 The magnetization hysteresis loops $M(H)$ recorded for various temperatures.

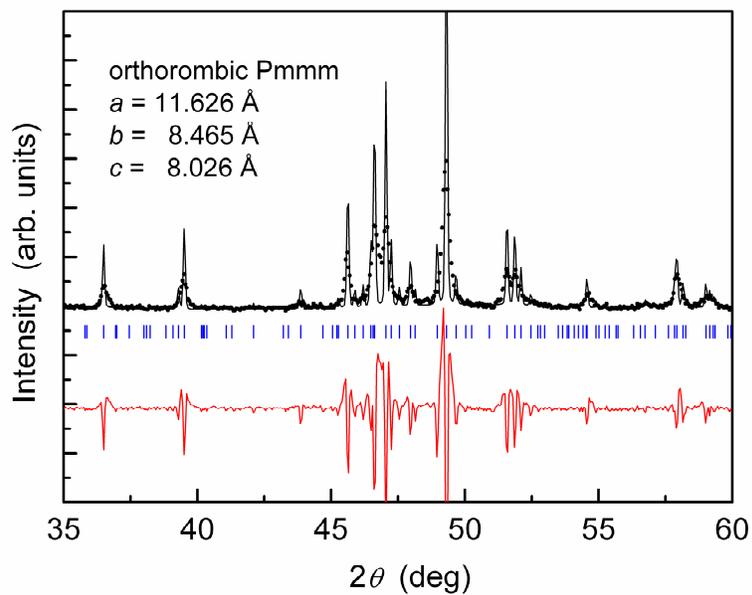

Fig.1

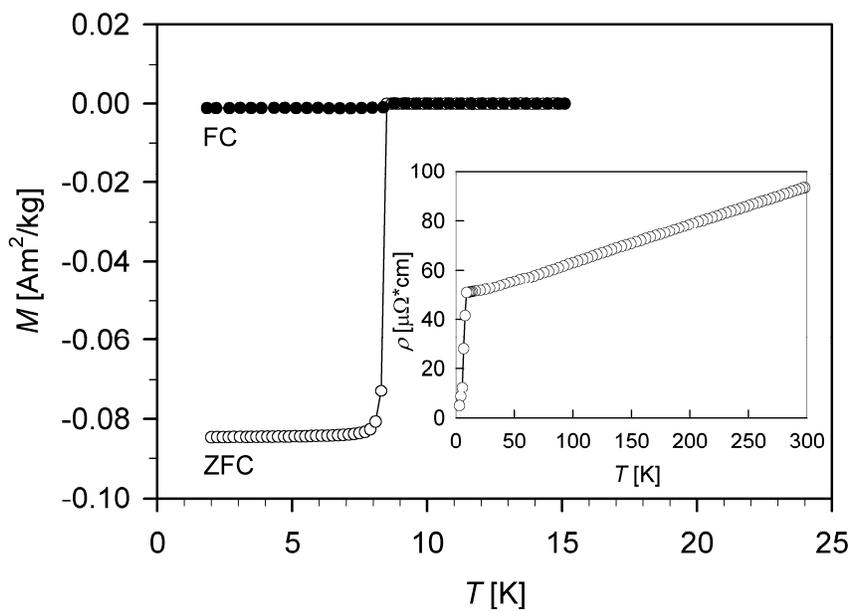

Fig.2

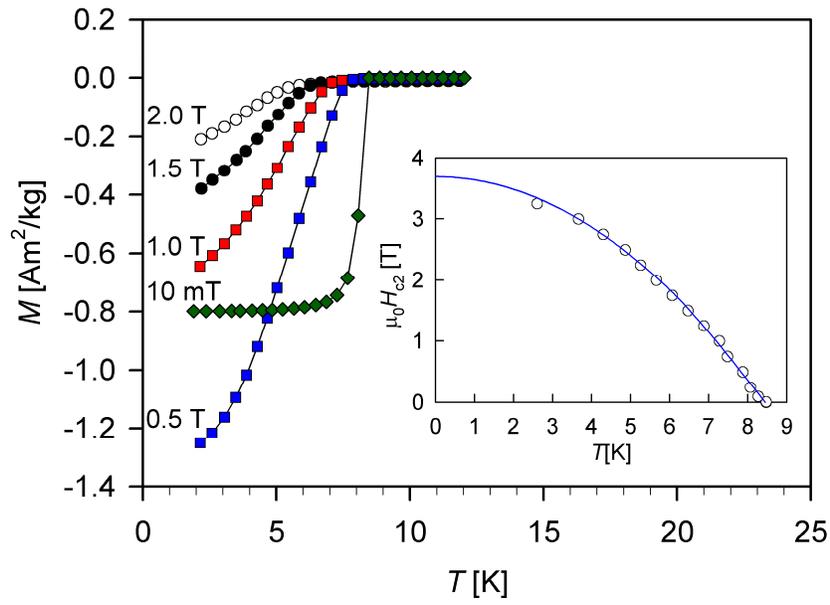

Fig.3

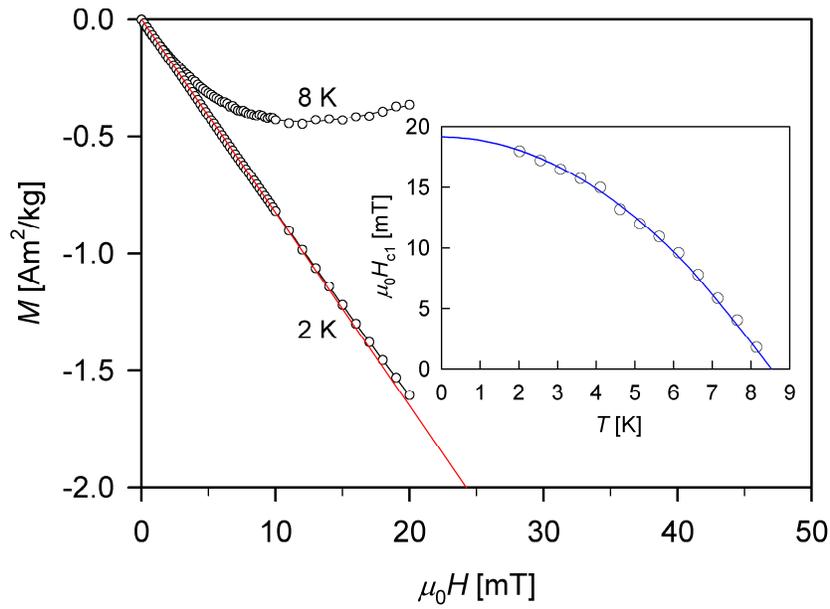

Fig.4

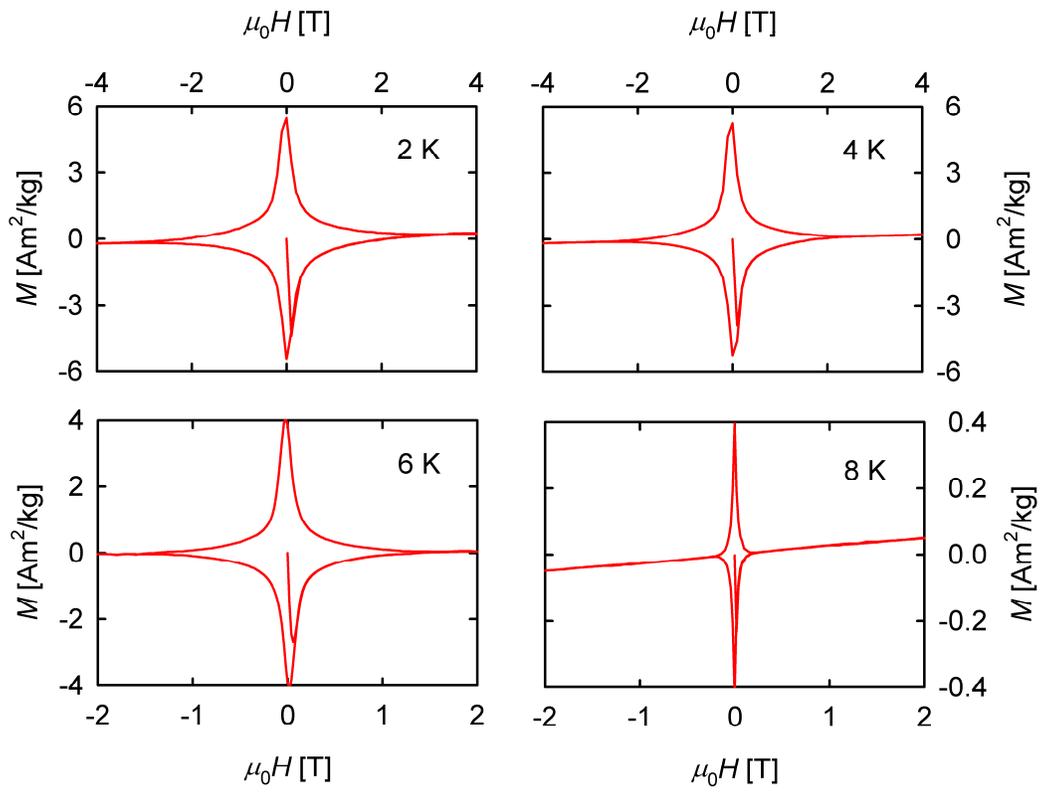

Fig.5